\documentclass{article}

\usepackage{PRIMEarxiv}

\usepackage[utf8]{inputenc} % allow utf-8 input
\usepackage[T1]{fontenc}    % use 8-bit T1 fonts
\usepackage{hyperref}       % hyperlinks
\usepackage{url}            % simple URL typesetting
\usepackage{booktabs}       % professional-quality tables
\usepackage{amsfonts}       % blackboard math symbols
\usepackage{nicefrac}       % compact symbols for 1/2, etc.
\usepackage{microtype}      % microtypography
\usepackage{lipsum}
\usepackage{amsmath, amssymb}
\usepackage{fancyhdr}       % header
\usepackage{graphicx}       % graphics
\usepackage{xcolor}
\usepackage{algorithmic}
\usepackage{algorithm}

\newcommand{\ns}[1]{{\color{black}#1}}

\graphicspath{{media/}}     % organize your images and other figures under media/ folder

\title{Data-Driven Moving Horizon Estimation Using Bayesian Optimization
}

\author{
Qing Sun, Shuai Niu,  Minrui Fei
}

\begin{document}
\maketitle

\begin{abstract}
In this work, an innovative data-driven moving horizon state estimation is proposed for model dynamic-unknown systems based on Bayesian optimization. As long as the measurement data is received,  a locally linear dynamics model can be obtained from one Bayesian optimization-based offline learning framework. Herein, the learned model is continuously updated iteratively based on the actual observed data to approximate the actual system dynamic with the intent of minimizing the cost function of the moving horizon estimator until the desired performance is achieved. Meanwhile,  the characteristics of Bayesian optimization can guarantee the closest approximation of the learned model to the actual system dynamic. Thus,  one effective data-driven moving horizon estimator can be designed further on the basis of this learned model. Finally,  the efficiency of the proposed state estimation algorithm is demonstrated by several numerical simulations.
\end{abstract}

% keywords can be removed
\keywords{Moving horizon state estimation, data-driven, Bayesian optimization, model dynamic-unknown system, learning framework.}

\section{Introduction}
{I}{n} practical applications, numerous physical models remain unknown, often categorized as black-box models. The dynamic model and noise distribution of such systems typically exhibit uncertainty or inaccuracy. Consequently, attaining an accurate reconstruction of the system state becomes challenging when faced with significant model uncertainty. Standard state estimation methods, including the Kalman filter (KF), Particle filter (PF), extended Kalman filter (EKF), and moving horizon estimation (MHE)~\cite{rawlings2006particle,alessandri2010advances,battistelli2018distributed,allan2019moving,schiller2023lyapunov}, struggle to achieve precise system state reconstruction under substantial model uncertainty. While H-infinity($H_{\infty}$) can handle scenarios where certain parameters of the model are unknown, it is inadequate for completely unknown systems~\cite{simon2006optimal}. Moreover, the complexity of practical systems introduces an additional challenge in accurately modeling their true dynamics. As commonly acknowledged, modeling more intricate systems often leads to notable uncertainties.

To address the practical limitations of conventional filtering methods caused by uncertainties in system models, a data-driven filter reconstruction approach based on Bayesian optimization (BO) strategy is proposed. The key advantage of BO lies in its ability to bypass the need for costly modeling of the true system dynamics. 
Instead, it employs a response surface methodology~\cite{khuri2010response} to link explanatory variables to resonse ones to fit a surrogate model, on which a global optimization is solved to accurately evaluate the objective function.
In other words, BO enables the optimization of a single dynamics model with respect to the design objective function, leading to the acquisition of a data-driven model.

In previous studies, BO techniques have primarily been proposed for designing data-driven controller parameters. For instance, Calandra \emph{et al}. utilized Bayesian optimization techniques to design and control robot motion gates~\cite{calandra2016bayesian}. In reference~\cite{rokonuzzaman2021customisable}, a new model predictive control (MPC) based on parameterized cost function is designed for longitudinal control of vehicles. BO has also been explored for parameter selection in controllers such as PID, LQR, and MPC to address model uncertainty, as investigated in~\cite{piga2019performance,marco2016automatic,trimpe2014self,sorourifar2021data}. 
However, it has been argued by Bansal et al. that solely adjusting controller parameters or the penalty matrix may not adequately meet the requirements when faced with significant uncertainty in the system model~\cite{bansal2017goal}. 
Consequently, they have proposed an active learning framework for dynamical models with the objective of identifying the best-performing model and employing it in controller construction. Nevertheless, this design concept precisely aligns with the objective of addressing filtering challenges in complex or dynamically-unknown systems, an area that has remained largely unexplored in earlier investigations. 
However, there is limited research on the application of Bayesian optimization in state estimation. The Moving Horizon Estimation (MHE) method heavily relies on accurate models, and it still has limitations when it comes to non-linear and perturbed systems~\cite{alessandri2010advances}.  
In reference~\cite{muntwiler2022learning}, a MHE method based on disciplined parametrized programming has been proposed, with the primary contribution being the online optimization of system parameters in Moving Horizon Estimation (MHE) to achieve enhanced state estimation results. Jin et al has provided a comprehensive review of past studies on data-driven state estimation and discusses future trends in state estimation. Among these trends, Bayesian optimization is identified as a highly promising method with significant potential~\cite{jin2021new}.

In this paper, we propose a data-driven MHE method based on BO. 
The method is primarily designed to address the case where no prior information about the system dynamics is available. 
By employing continuous learning, the proposed approach takes into account the presence of measurement errors and noise interference in the context of MHE, utilizing continuous learning in the absence of prior knowledge about the system dynamics model. Unlike traditional state estimation methods that heavily rely on model-driven approaches, the proposed strategy formulates models describing data relationships through Bayesian optimization-offline learning, eliminating the need for precise model acquisition and minimizing information loss during the fitting process.
For completely unknown systems, the optimization-based learning framework enables the derivation of a locally linear dynamics model. The primary objective is to minimize the cost function of the moving horizon estimator, aiming to approximate the maximum filtering accuracy rather than achieving complete reconstruction of the actual system. By avoiding the computational burden and errors associated with nonlinear system computations, this approach evaluates overall performance and mitigates the losses incurred by linearized models to a certain extent.

The paper is structured as follows: The problem description of the nonlinear system is presented in Section~\ref{sec:problem_formulation}, where the fundamental theory of Moving Horizon Estimation (MHE) for this system and the treatment of unknown system parameters are discussed. Section III provides a comprehensive explanation of the general procedure of BO. The integration of BO and MHE is discussed in detail in Section IV, outlining the specific intricacies of the proposed BOMHE algorithm. In Section V, two numerical experiments are conducted to demonstrate and analyze the performance of the algorithm.\\

\noindent\textbf{Notation}:
For the data sequence $x_0,x_1,...,x_T$, $T \in \mathbb{N}$, we write $(x_k)_{k=0}^T$.  For a vector $x \in \mathbb{R}^n$ and a symmetric and positive definite matrix $M \in \mathbb{R}^{n \times n}$, $|| x ||_M^2 := x^\top M x$ is a weighted Euclidean norm, where $\cdot^\top$ denotes transposition. 
Moreover, $\operatorname{diag}(x)$ stands for the $(n \times n)$-diagonal matrix~$L$ with entries $L_{ii} = x_i$, $i \in [1:n] := [1,n] \cap \mathbb{Z}$. $\overline{x}_{t-N}$ denotes the prediction of the $x_{t-N}$ at the previous time instant t-1.
$GP$ denotes the Gaussian process: $J \sim \mathcal{G} \mathcal{P}\left(m(\theta), k\left(\theta, \theta^{\prime}\right)\right)$,the function $J$ is distributed as a Gaussian process with mean function $m(\theta)$ and covariance function $k(\theta, \theta^{\prime})$.$K$ is the n×n covariance gram matrix $K(\theta, \theta)$.

\section{Problem formulation}
\label{sec:problem_formulation}

\ns{A dynamic system described by the discrete-time equations is considered as follows:}
\begin{equation}\label{equ:1}
\begin{split}
    x_{k+1} &= f\left(x_{k},  u_{k}\right)+w_{k},\\
    y_{k} &= h\left(x_{k}\right)+v_{k}
\end{split}
\end{equation}
\ns{for $k \in \{0, 1, ..., T\}$ with functions $f: \mathbb{R}^{n_x} \times \mathbb{R}^{n_u} \rightarrow \mathbb{R}^{n_x}$ and $h: \mathbb{R}^{n_x} \rightarrow \mathbb{R}^{n_y}$ and initial state $x_0 \in \mathbb{R}^{n_x}$. 
%where the functions $f(\cdot)$ and $h(\cdot)$ denote the state equation and observation equation, respectively. 
\ns{The state, control and measurement output of the system at time $k$ are denoted by $x_k\in\mathbb{R}^{n_x}$% is the system state
, $u_k\in\mathbb{R}^{n_u}$% is a known input
, and $y_k\in\mathbb{R}^{n_y}$% is a output of measurements
, r}espectively. 
The vectors $w_k \in \mathbb{R}^{n_w}$ and $v_k \in \mathbb{R}^{n_v}$ correspond to the process noise and measurement noise, respectively. Either $n_w = n_x$ and $n_v = n_y$.} 
Note that the specific distribution of the noise in this scenario is unknown.

\ns{This paper aims to develop} a data-driven estimator by leveraging Bayesian optimization of Gaussian processes in conjunction with \ns{MHE}. The primary objective is to solve the state estimation problem, which involves determining the true state $(x_k)_{k=0}^T$ of the system based on noisy measurements $(y_k)_{k=0}^T$. % obtained from the measurement equation (1) that accounts for system and model noise. 
MHE, a variant derived from total information estimation, addresses the computational complexity associated with the entire information set over time. Instead, MHE focuses on the computational cost within a finite time window of \ns{$N+1$} measurements.

A sliding window $[t-N: t]$ is established, where $(x_k)_{k=t-N}^t$ represents the \ns{$N+1$} most recent values of the state sequence at the current \ns{time~$t$}. %$T$. 
At each step, an estimate is generated for a truncated sequence $(\hat{x}_k)_{k=t-N}^t$ of length \ns{$N+1$}. Considering the unknown statistical properties of the perturbations, we employ %ordinary 
least squares to derive the estimates with the cost function following: 
\ns{\begin{equation}\label{equ:2}
\begin{split}
J_t &=\sum_{k=t-N}^{t}||y_{k}-h\left(\hat{x}_{k|t}\right)||_{R}^{2}+\\
&\sum_{k=t-N}^{t-1}||\hat{x}_{k+1|t} -f\left(\hat{x}_{k|t},  u_{k}\right)||_{Q}^{2}+\Gamma_{t-N}(\hat{x}_{t-N|t}),\\
&\Gamma_{t-N}(\hat{x}_{t-N|t})=||\hat{x}_{t-N|t}-\overline{x}_{t-N}||^2_{P^-_{t-N}}
\end{split}
\end{equation}}
subject to \ns{system~\eqref{equ:1}, $k \in [0:T]$,
where $\hat{x}_{k|t}$ denotes the estimate of $x_k$ to be made at any stage t, $||y_{k}-h\left(\hat{x}_{k|t}\right)||_{R}^{2}$ denotes the (weighted) distance between the expected output and the actual measurement (equal to $\| v_k \|_R$) and $||\hat{x}_{k+1|t} -f\left(\hat{x}_{k|t},  u_{k}\right)||_{Q}^{2}$ the variation of the system itself (equal to $\| w_k \|_Q)$. In the following, we use positive definite and diagonal weighting matrices~$Q$ and~$R$.} % are  that can be set artificially. 
Where $P_{t-N}$ is a positive definite weight matrix quantifying the confidence on the prediction $\overline{x}_{t-N}$.
The third term $\Gamma_{t-N}(\cdot)$ is referred to as the arrival cost and serves the purpose \ns{of summarizing past data before time $t-N$}. $\Gamma_{t-N}(\cdot)$ summarizes the impact of measurement data $(y_k)_{k=0}^{t-N-1}$ on state $\hat{x}_{t-N|t}$. \ns{The selection of the arrival cost $\Gamma_{t-N}(\cdot)$ is a critical factor in determining the behavior and efficacy of an MHE scheme~\cite{alessandri2008moving}.}

It follows from (2) that the main problem in designing this class of MHE is to model the unknown system dynamics in (1). As done in the EKF,  the linearized system matrix of a nonlinear system is usually obtained using a first-order Taylor expansion. Here we take the linear time-invariant system matrix \ns{$(A, B, C^\top) \in \mathbb{R}^{n_x \times (n_x + n_u + n_y)}$} as an approximation to the real system. 
The system matrix is then parametrised by $\theta \in \Theta \subseteq \mathbb{R}^d$ %as $\theta:=(A, B, C^\top)$ 
\ns{with $d = n_x \cdot (n_x+n_u+n_y)$, where} $\Theta$ is the domain of the $\theta$. The active learning process takes a set of system matrices $\theta$ at each time k given the current system matrix $\theta$ and the initial value of the state $x_0$. Unlike the traditional approach,  in each active learning process,  a set of system matrices $\theta$ is taken,  and we denote the sequence of state estimates obtained during the optimization of this linear system by using $J(\theta)$:
\begin{equation}\label{equ:4}
\begin{split}
J(\theta)&= \sum_{k=t-N}^{t}||y_{k}-h\left(\hat{x}_{k|t}\right)||_{R}^{2}\\
&+\sum_{k=t-N}^{t-1}||\hat{x}_{k+1|t} -f\left(\hat{x}_{k|t},  u_{k}\right)||_{Q}^{2}+\Gamma_{t-N}(\hat{x}_{t-N|t}),
\end{split}
\end{equation}
subject to:
\begin{equation}\label{equ:5}
\begin{split}
x_{k+1} &= Ax_k+Bu_k+w_k,\\
y_{k} &= Cx_k+v_{k}.\\
\end{split}
\end{equation}
%where \ns{$\theta = (A,B,C^\top) \in \mathbb{R}^{n_x \times (n_x + n_u + n_y)}$}.  %$C\in\mathbb{R}^{n_y\times n_u}$.

The advantage is that we do not need to explicitly compute the linearized matrix of the real system each time but use $\theta$ instead each time when $\theta$ changes (i.e.,  taking different system matrices $\theta:=(A, B, C^\top)$,  each time we can get the corresponding sequence of state estimates $J ( \theta)$,  we aim to find an optimal set of $\theta^*:=(A^*,  B^*,  C^{\top*})$  so that the resulting estimates are closest to the state values of the real system. Thus, the process of finding the optimal parameters can be described as follows:
\begin{equation}\label{equ:7}
\theta^{*}=\operatorname{argmin}_{\theta \epsilon \Theta} J(\theta).
\end{equation}

In this paper,  the linear time-invariant model is chosen to replace the actual parameters of the system. The advantage is that the computational effort required for physical modeling can be effectively reduced through a dynamic learning process from the data, considering the uncertainty created by noise. 
For the case where the actual system is a linear model, the active learning framework is very accurate and can almost completely restore the actual system parameters.
\ns{For complex nonlinear systems, the typical approach is to perform linearization approximation. However, the obtained parameters of the linear system may exhibit strong coupling or temporal variations},  performance may be lost at some time through this linearisation. However,  the simulation model \ns{resulting} from this framework is overall close to the real performance of the system. Since the actual system with \ns{$f(\cdot), h(\cdot)$} is usually unknown,  so for the specific shape of the cost function (2) in the MHE is also unknown; it is challenging to get the actual shape of the cost function,  which is taken empirically in many experiments,  while another advantage in the framework of this paper is that the optimal value of the cost function can be \ns{evaluated} through Bayesian optimization, which only generates less computational effort.

\section{Bayesian optimization}
Bayesian optimization is a global optimization method based on a surrogate model. For systems with unknown objective functions and high computational complexity, \ns{Bayesian optimization has demonstrated superior performance in such cases.}
The posterior distribution of the objective function is estimated based on the data, and then the next combination of hyperparameters to be sampled is selected based on the distribution of the posterior distribution\cite{beckers2019closed,chakrabarty2021safe,lederer2019uniform}. \ns{In this paper, the desired hyperparameter is one of the linearized system parameter sets $\theta$ of the system.}
It learns the objective function and determines the parameters contributing to the global maximum. The form of the objective function is learned by assuming an acquisition function based on a prior distribution. Each time the objective function is tested with a new sampling point,  this information is used to update the prior distribution of the objective function. The final test of the minimum value point given by the posterior distribution\cite{bansal2017goal}.

In Bayesian optimization,  we first need an efficient way to model the distribution of the objective function. Therefore,  we usually based on the unknown objective function to create a dataset $D=\{\theta,  J(\theta)\}$ with respect to the parameters $\theta$,  where it is important to note that we still do not know the exact shape of the objective function $J(\theta)$ and cannot obtain information about the gradient of the model during training. We want to model $J(\theta)$ with a confidence distribution to find the optimal solution,  so we set an approximate model $\hat{J}(\theta) $ to obtain the true value of $J(\theta)$ corresponding to $\theta$,  that is, $\hat{J}(.)$ is a mapping from $\theta$ to $\hat{J}(\theta)$ likes $J(.)$, so the approximate model $\hat{J}(\theta)$ is used in equation (6) instead of $J(\theta)$.
\begin{equation}\label{equ:8}
\begin{split}
\theta^{*}=\operatorname{argmin}_{\theta \in \Theta}{\hat{J}}(\theta).
\end{split}
\end{equation}

In this paper,  it is only the approximate model that needs to be evaluated according to $\hat{J}(\cdot) $,  not the actual objective function,  and when Bayesian optimization is performed on the actual objective function according to $\hat{J}(\cdot) $ obtains a new set of parameters $\theta$ before it is finally evaluated in the actual objective function $J(\theta)$. A probabilistic model is used in Bayesian optimization. The advantage of using a probabilistic model is that it can directly model observations with noise and completely take into account the uncertainty of the system itself without having to consider the distribution in which the noise is located specifically. This has the advantage that the probabilistic model used is more robust when the effect of model error is present.
There are many choices for $\hat{J}(\cdot) $. Usually, one can use random forest or Gaussian process as the approximate probabilistic model\ns{\cite{bansal2017goal}.} This paper uses the Gaussian process as the optimized probabilistic model.

\subsection{Gaussian Process}

In (5),  the specific information of $J(\theta)$ is unknown. Therefore,  the $\hat{J}(\cdot)$ generated by the Gaussian process is used to approximate the prior distribution of $J(\theta)$. The Gaussian Process is a Bayesian nonparametric model of the joint distribution of random variables over a continuous domain,  where the union of any finite number of distinct random variables over a continuous domain has a joint Gaussian distribution~\cite{seeger2004gaussian}. We can treat any function as a vector of infinite length,  where each entry in that function is a function value $J(\theta)$ with specified input $\theta$. Thus assuming that the different values of the loss function are associated with $\theta$ $\hat{J}(\cdot) $ are random variables and that these random variables satisfy the joint Gaussian distribution assumptions above,  then the Gaussian process with respect to that random variable is a distribution over $\hat{J}(\theta) \sim G P\left(m(\theta),  k\left(\theta,  \theta^{\prime}\right)\right)$ where  Gaussian process is determined when the mean $m(\theta)$ and the covariance $k(\theta, \theta^{\prime})$ are uniquely given. The mean and covariance functions of  the Gaussian process are usually described as follows:
\begin{equation}\label{equ:9}
\begin{split}
m(\theta)&=E[\hat{J}(\theta)],\\
k(\theta, \theta^{\prime})&=\operatorname{cov}\left(\hat{J}(\theta), \hat{J}\left(\theta^{\prime}\right)\right).
\end{split}
\end{equation}

With the above equation,  we obtain the hyperparameters that can fit any model. In many studies,  the mean function is usually set to zero,  but if there is a priori knowledge about the dynamics or training data of the system to be simulated,  for example,  if the dynamics or training data have a linear trend, then choosing a non-zero mean function will not only speed up the convergence of the model,  but also reduce the uncertainty of the model,  but will increase the computational burden accordingly. 
\ns{Usually}, %In most cases,  
the choice of covariance function is  \ns{the more challenging task.} %usually more concerning. 
In machine learning,  the covariance function is usually called the kernel function,  through which the features of the model are extracted. The kernel function achieves the prediction of unknown points by collecting the relationship between different points and reflecting them in the position of the sample afterward. The choice of kernel function has a great influence on the training effect of the system ~\cite{klenske2015gaussian}. The commonly used kernel functions are Periodic kernel (Periodic),  Linear kernel (Linear),  Square exponential kernel (SE), etc. In this paper we use a square exponential kernel,  about the SE kernel between pairs of random variables can be described as:
\begin{equation}\label{equ:10}
\begin{split}
k_{J}\left(\theta,  \theta^{\prime}\right)=\sigma_{J}^{2} \exp \left(-\frac{1}{2}\left(\theta-\theta^{\prime}\right)^{T} \boldsymbol{\Lambda}^{-1}\left(\theta-\theta^{\prime}\right)\right),
\end{split}
\end{equation}
where $\sigma_{J}^{2}$ is the variance of the function $\hat{J}(\cdot)$,  $\boldsymbol{\Lambda}=diag\left[l_{1}^{2},  l_{2}^{2}, \ldots,  l_{D}^{2}\right]$, \ns{where $l_{i}$ is a vector with positive values,the $l_{i}$ interprets as the
characteristic length scales~\cite{osborne2009gaussian}.} For convenience,  the uniform characteristic length scale is sometimes taken. These are adjustable hyperparameters. Therefore in order to maximize the probability of $\hat{J}(\cdot)$ arising under these two hyperparameters,  the optimal parameter~\cite{seeger2004gaussian} is found by maximizing the marginal log-likelihood (MarginalLog-likelihood).
For the currently available input data pairs $(\theta_i, \hat{J}(\theta_i))_{i=1, }^{i=n}$,  the joint distribution of these data points $[\hat{J}(\theta_1), \hat{J}(\theta_2), ..., \hat{J}(\theta_i)]_{i=1, }^{i=n}$ needs to satisfy the multidimensional Gaussian distribution according to the definition of Gaussian process as follows:
\begin{equation}\label{equ:11}
\begin{split}
\left[\hat{J}\left(\theta_{1}\right),  \hat{J}\left(\theta_{2}\right),  \ldots, \hat{J}\left(\theta_{n}\right)\right]^{T} \sim\mathcal{N}(\mu,  K),
\end{split}
\end{equation}
where $\mu=\left[m\left(\theta_{1}\right),  m\left(\theta_{2}\right),  \ldots,  m\left(\theta_{n}\right)\right]^{T}$ is the mean vector, $K$ is the kernel matrix with $K_{ij} =k(\theta_i, \theta_j)$
Then the joint distribution of function values $\hat{J}(\theta^{\prime})$ and past known points $(\theta, \hat{J}(\theta))$ for any point $\theta^{\prime}$ is
\begin{equation}\label{equ:12}
\begin{split}
\left[\begin{array}{l}
\hat{J}(\Theta) \\
\hat{J}\left(\theta^{\prime}\right)
\end{array}\right] \sim \mathcal{N}\left(m\left(\theta^{\prime}\right), \left[\begin{array}{ll}
K(\Theta, \Theta) & K\left(\Theta,  \mathbf\theta^{\prime}\right) \\
K\left(\theta^{\prime},  \Theta\right) & K\left(\theta^{\prime},  \theta^{\prime}\right)
\end{array}\right]\right).
\end{split}
\end{equation}
The distribution of the function value $\hat{J}(\theta^*)$) can be predicted by GP.
\begin{equation}\label{equ:13}
\begin{split}
&m\left(\theta^{\prime}\right)=K\left(\Theta,  \theta^{\prime}\right) K^{-1} \textbf{J},\\
&\sigma^{2}\left(\theta^{\prime}\right)=\boldsymbol{k}\left(\theta^{\prime},  \theta^{\prime}\right)-K\left(\Theta,  \theta^{\prime}\right) K^{-1} K\left(\Theta,  \theta^{\prime}\right)^{T},
\end{split}
\end{equation}
where $\Theta=\{\theta_{1}, \theta_{2}, ..., \theta_{n}\}$ denotes the set of past known points, $\textbf{J}=\{\hat{J}\left(\theta_{1}\right),  \hat{J}\left(\theta_{2}\right), \ldots, \hat{J}\left(\theta_{n}\right)\}$denotes the set of  estimated value, $K(\Theta, \theta^{\prime})=[k(\theta_{1}, \theta^{\prime}), ..., k(\theta_{n}, \theta^{\prime})]$ contains the information between the current best point and the other points, $m(\theta^{\prime})$ is the expected value of the prediction at that point $\left(\theta^{\prime}, \hat{J}\left(\theta^{\prime}\right)\right)$,  and $\sigma^{2}(\theta^{\prime})$ denotes the uncertainty of the estimate at that point.

\subsection{\ns{Acquisition} %re 
function}\label{sec:ei}

After building the initial GP model,  the next step is to find the sampling points for each GP update. The most straightforward approach is to sample some arbitrary points at all intervals,  but this results in a great computational effort and waste at sampling points with high uncertainty. Therefore,  a common approach is to guide the sampling point selection strategy based on the posterior model~\cite{shahriari2015taking}. Usually,  the strategies that can be chosen are probability of improvement(PI), expected improvement(EI), upper confidence bound(UCB), predictive entropy search(PES), etc. There are also strategies that use multiple acquisition functions. There are also methods that use a combination of multiple acquisition functions to perform selection,  such as entropy search portfolio (ESP). In GP,  it is usually necessary to consider the tradeoff between mean and variance. Choosing points with larger mean will help us better understand the true shape of the unknown function while choosing points with larger variance shows a greater exploration for less informative points,  so the collection function needs to be set to help avoid producing local optima or non-convergence. In this paper,  we choose EI as the search strategy. From the previous section,  we know that the function value $\hat{J}(\theta)$ at a point $\theta$ can be regarded as a normal random variable with mean $m(\theta)$ and variance $\sigma^2(\theta)$. The utility function of EI  is set to be as follows:
\begin{equation}
\mathrm{EI}(\mathbf{\theta})=\mathbb{E} \max \left(\hat{J}(\mathbf{\theta})-\hat{J}\left(\mathbf{\theta}^{+}\right), 0\right)
\end{equation} 
where $\hat{J}(\theta) \sim \mathcal{N}\left(\mu,  \sigma^{2}\right)$,$\hat{J}(\theta^+)$  denotes the value of the best sample so far.

\begin{equation}
\begin{aligned}
\mathrm{EI}(\theta) & = \begin{cases}\left(\mu(\theta)-\hat{J}\left(\theta^{+}\right)-\xi\right) \Phi(Z)& \text { if } \sigma(\theta)>0\\
+\sigma(\theta) \phi(Z)  \\
0 & \text { if } \sigma(\theta)=0\end{cases} \\
\mathrm{Z}  &= \frac{\mu(\theta)-\hat{J}\left(\theta^{+}\right)-\xi}{\sigma(\theta)} 
\end{aligned}
\end{equation}
where $\mu(\theta)$ and $\sigma(\theta)$ are the mean and the standard deviation of the GP posterior predictive at $\theta$, respectively.
where $\phi(.) $ is the probability density function (PDF), $\Phi(.) $ is the cumulative distribution function (CDF). Thus the next best sampling point $\theta_{next}$ can be derived by maximizing the expectation function as follows:
\begin{equation}\label{equ:17}
\begin{split}
\theta_{next}=\operatorname{argmax_{\theta}} {EI(\theta)},
\end{split}
\end{equation}

\section{Data-driven filter via \ns{Bayesian} optimization}\label{sec:BOMHE}

In this section, the active learning framework of MHE will be explicitly constructed and proposed for nonlinear systems of more complex form with uncertain model knowledge. The core idea of this method is that a novel framework based on BO is utilized to obtain the optimal model dynamics by maximizing the cost function within MHE filter framework.

The detailed steps are given as follows. Firstly,  $\theta$ is used instead of the actual system parameters, and the process of finding the optimal alternative parameters is represented by equation (5). At this time,  for a given parameter $\theta$,  there is always the output value of the corresponding objective function $J(\theta)$. However,  at this time,  the specific form of the objective function is unknown (that is, the mapping relationship of $\theta$ on $J(\theta)$ cannot be obtained),  so we cannot obtain the gradient information during the training of the model,  and at this time the problem is transformed into a black-box optimization problem. Therefore,  Bayesian optimization is considered,  and the probabilistic model of the objective function is built using GP to achieve the optimization problem of (7).
In this algorithm, linearized system parameters are used as optimization variables, and the arrival cost can be expressed as
\begin{equation}\label{equ:18}
\begin{split}
&\Gamma_{k}=||\hat{x}_{k|t}-\overline{x}_{k}||^2_{P^-_{k}},
\end{split}
\end{equation}
where the matrix ${P_k}\in\mathbb{R}^{n_x\times n_x}$ can be obtained by solving the Riccati equation derived from the standard Kalman filter as follows:
\begin{equation}
\begin{split}
&P_{k+1}^{-}=A_{k}P_{k}^{+} A_{k}^{T}+Q\\
&P_{k}^{+} = (I-K_{k}C_{k})P_{k}^{-}\\
&K_{k}=P_{k}^{-}C_{k}^{T}(C_{k}P_{k}^{-}C_{k}^{T}+R)^{-}
\end{split}
\nonumber
\end{equation}
\begin{equation}\label{equ:19}
\begin{split}
P_{k+1}^{-}&=A_{k}P_{k}^{-} A_{k}^{T}-A_{k} P_{k}^{-} C_{k}^{T}(C_{k} P_{k}^{-} C_{k}^{T}+R)^{-1} C_{k} P_{k}^{-} A_{k}^{T}\\
&+Q.
\end{split}
\end{equation}
Therefore, when constructing a filter, first giving the initial value of the state estimate $\hat{x}_0$,  any parameter $\theta:=(A, B, C^\top)$,  the initial value of the arrival cost coefficient $P_0$,  the sequence of observations of the real system $(y_{k})_{k=0}^T$,  MHE by equation (3),  each optimization takes only the first value of the obtained sequence of state estimates $(\hat{x}_{k}^{\text {mhe}})_{k=t-N}^t$as the valid state estimate and uses it as the initial value for the next optimization,  in each MHE optimization to obtain the cost function $J(\theta)$ equation (6) until all the state estimates $(\hat{x}_{k}^{\text {mhe}})_{k=0}^T$ are obtained.

Using this equation to measure the difference between the state estimates obtained by the system under the action of the parameter $\theta$ after MHE and the observed values and the evolution of the system itself,  then $J(\theta)$ at this point contains all the information obtained within the window. Hence,  the data set $\mathcal{D}=\{\theta_i,  J(\theta_i)\}$ is used in the active learning process for Bayesian optimization. A complete MHE process is required in each step of Bayesian optimization,  where the $(\theta_i,  J(\theta_i))$ obtained each time is used to update the GP,  and then the next obtained by solving the minimization Acquire function (14) for the next parameter point to be sampled $\theta=\theta_{next}$,  and repeat this process until the maximum number of iterations is reached. The exact procedure is summarized in the BOMHE algorithm.
\begin{algorithm}[H]
\caption{BOMHE algorithm}\label{alg:alg1}
\begin{algorithmic}[1]
\STATE {\textsc{INPUT}} $\mathcal{D}=\{\theta_0, J(\theta_0)\}, P_0, (y_k)_{k=0}^T, \hat{x}_0$
\STATE Initialize GP with $\mathcal{D}$
\STATE for {$i = 1 \to Maxiter$} do
\STATE \hspace{0.5cm}Find $\theta_{next}$ using (14)
\STATE \hspace{0.5cm}$\theta_i \leftarrow \theta_{next},  X_{\theta_i}^{mhe}=\{\}$
\STATE \hspace{0.5cm}for {$t = N \to T$} do
\STATE \hspace{1cm}Given$(y_k)_{k=t-N}^{t-1}, x_0, P_0, \theta_i:=(A_i, B_i,C^\top_i)$
\STATE \hspace{1cm}Compute $(x_k^{mhe})_{k=t-N}^t, J^{'}(\theta_i)$ using(7)
\STATE \hspace{1cm}Update $P_{t-N+1}$
\STATE \hspace{1cm}$P_0\leftarrow P_{t-N+1}$
\STATE \hspace{1cm}$J(\theta_i)=J(\theta_i)+J^{'}(\theta_i)$
\STATE \hspace{1cm}$X_{\theta_i}^{mhe}\leftarrow (\hat{x}_{k}^{\text {mhe}})_{k=0}^T$
\STATE \hspace{0.5cm}end for
\STATE \hspace{0.5cm}return $\theta_i, J(\theta_i), X_{\theta_i}^{mhe}$
\STATE \hspace{0.5cm}Using $(\theta_i, J(\theta_i))$ update GP and $\mathcal{D}$
\STATE end for
\end{algorithmic}
\label{alg1}
\end{algorithm}
        
It should be noted that in this optimization process,  we do not directly use the MHE objective function containing the parameters of the nonlinear system as the object of the action of the surrogate model generated by GP. However,  we use $J(\theta)$ to contain its output values under the action of the parameter $\theta$,  so there should be an additional choice about the specific form of $J(\theta)$. Due to the setting of the acquisition function,  the process of selecting the parameter $\theta$ is not entirely random,  it strategically goes to acquire the next parameter value to learn the shape of the whole function by evaluating the characteristics of the past dataset. The search behavior of this method ensures to some extent the efficiency of the training to obtain the optimized global minimum faster. Because of this,  this active learning framework will converge faster if some information related to the actual system parameters can be obtained a priori and used to initialize in GP or probe in the specified domain.

\section{Numerical simulations}\label{sec:2}
In this section,  we illustrate the effectiveness of this BOMHE algorithm through simulation examples.
\subsection{Leak estimation}\label{sec:3}
\begin{figure*}[!t]
    \centering
    \includegraphics[width=\textwidth]{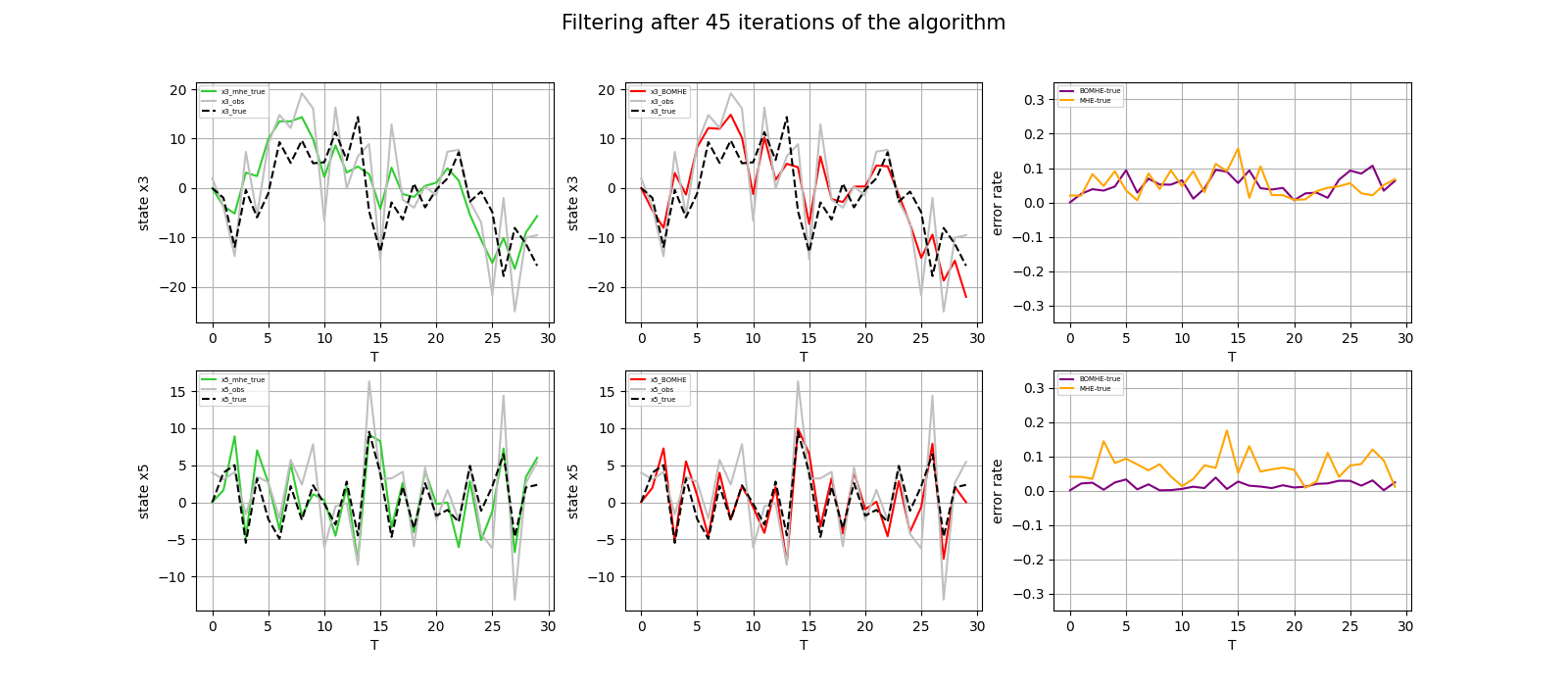}
    \caption{The simulation results provide the outcomes after the algorithm iterates for 45 times. In the figure, the red solid line represents the estimate obtained through the BOMHE algorithm, while the green dashed line represents the estimate computed using the conventional MHE. The gray solid line depicts the observed values of the system under the influence of noise. The black dashed line illustrates the actual state of the system. The purple and orange solid lines respectively denote the discrepancies between the estimates obtained by the BOMHE algorithm and the conventional MHE algorithm, and the true state of the system.}
    \label{fig:2}
\end{figure*}

We first consider the state estimation process of the leak detection device described on ~\cite{rao2002constrained}.
\[x_{k+1}=\begin{bmatrix}
0.89168 & 0 & 0 & 0 & 1.0 \\
0.10832 & 0.90518 & 0 & 0.04306 & 0 \\
0 & 0.09482 & 0.89524 & 0 & 0 \\
0 & 0 & 0.10476 & 0.89235 & 0 \\
0 & 0 & 0 & 0 & 0
\end{bmatrix}
x_{k}\]
\[+\begin{bmatrix}
-1 & 0 & 0 & 0 & 0 \\
0 & -1 & 0 & 0 & 0 \\
0 & 0 & -1 & 0 & 0 \\
0 & 0 & 0 & -1 & 0 \\
0 & 0 & 0 & 0 & 1
\end{bmatrix}
w_k\]

\[y_{k}=\begin{bmatrix}
1 & 0 & 0 & 0 & 0 \\
0 & 1 & 0 & 0 & 0 \\
0 & 0 & 1 & 0 & 0 \\
0 & 0 & 0 & 1 & 0 \\
0 & 0 & 0 & 0 & 1
\end{bmatrix}
x_{k}+v_{k}\]
The state vector $x_k$ of the system includes the state values $(x_{k,1}, x_{k,2}, x_{k,3}, x_{k,4}, x_{k,5})$ at the locations to be detected,  where we assume that the real model has perturbations only at states $x_{k,3}$ and $x_{k,5}$,  so $w_k$ is a normally distributed random variable with covariance matrix $Q_w$,  and $v_k$ is a normally distributed random variable with covariance matrix $R_v$.
\begin{equation}
\begin{split}
&Q_w=diag[0, 0, 5, 0, 15]\\
&R_v=diag[8, 8, 8, 8, 4]
\end{split}
\nonumber
\end{equation}
Using equation (6) as the cost function for state estimation,  we assume that the dynamics of the system are unknown in the case of estimating the state values $x_3$ and $x_5$ using the BOMHE algorithm,  taking the estimated time-domain N=10. Since the location of the leakage is unknown in the state estimation,  the covariance matrices Q and R in the cost function are taken to be conservative,  respectively.
\begin{equation}
\begin{split}
&Q=diag[5, 5, 5, 5, 15]\\
&R=diag[8, 8, 8, 8, 4]
\end{split}
\nonumber
\end{equation}
Since our primary concern in this model is the state matrix,  let $\theta:=A_{\theta}$,  which corresponds to a model with five states,  of which a total of 9 parameters are learned in the BOMHE algorithm,  where each parameter corresponds to each non-zero numerical term in $\theta$. It is assumed here that there is no prior knowledge about the dynamics of the model,  so given the prior distribution of GPs,  $\theta$ is chosen to take random values,  and the boundary limit for each parameter is taken to be [0, 5]. The acquisition functions EI is used in the BOMHE algorithm in Eq. (14),  and in this framework,  we do not further optimize the EI capture function. For this linear model,  the BOMHE algorithm can search directly in the parameter space for the convergence parameter that minimizes the cost function,  which may not be the same as the actual system. However,  its filtering effect in that period should be the closest to the actual system. All observations within the observation period $T=90$ are taken. By the 45rd iteration of the algorithm,  a better solution has emerged. Since the model fitted by GP may be nonconvex,  EI may prefer to choose unexplored points for the next sampling point selection,  which has the disadvantage of generating unnecessary computational effort. Therefore,  Figure \ref{fig:2} performs the system simulated by the algorithm on the MHE at the 36th iteration of the algorithm.
Also,  to facilitate the evaluation of the effectiveness of BOMHE,  we use the mean absolute error (MAE) to give the filtering accuracy obtained by the MHE optimization as follows.
\begin{equation}
\begin{split}
M A E=\frac{1}{T} \sum_{k=1}^{k=T}\left(x_{k}-x_{k}^{m h e}\right),
\end{split}
\end{equation}
where $x_k$ denotes the real state value of the system at the current sampling time T and $x_{k}^{mhe}$ denotes the estimated value obtained by MHE. For this linear system,  the approximate model obtained by BOMHE does not completely reconstruct the actual system parameters. However,  the filtering performance is very similar to that of the actual system.

\begin{table}[!t]
\setlength{\abovecaptionskip}{0cm}
\setlength{\belowcaptionskip}{0.3cm} 
\caption{The leak estimation\label{tab:table1}}
\centering
\begin{tabular}{|c|c|c|}
\hline
ALGORITHM& MHE-TRUE &  BOMHE\\
\hline
MAE & 8.513 & 8.522\\
\hline
\end{tabular}
\end{table}

\subsection{Third-order heat transfer system}\label{sec:nonlinear}
We then consider a Third-order heat transfer system and test the performance of the algorithm on the nonlinear system as follows:
\begin{equation}
\begin{split}
\dot{x}_1 &= -k_1 \cdot (x_1 - T_{\text{env}}) + k_2 \cdot x_3 + k_u \cdot u +w_{x}\\
\dot{x}_2 &= 1 \\
\dot{x}_3 &= -k_3 \cdot (x_1 - T_{\text{env}})\\
y &= Cx + v_{y}\\
Q&=\operatorname{diag}[1, 1, 1],\\
R&=\operatorname{diag}[5, 5],
\end{split}
\nonumber
\end{equation}

\begin{figure*}[!t]
\centering
\includegraphics[width=\textwidth]{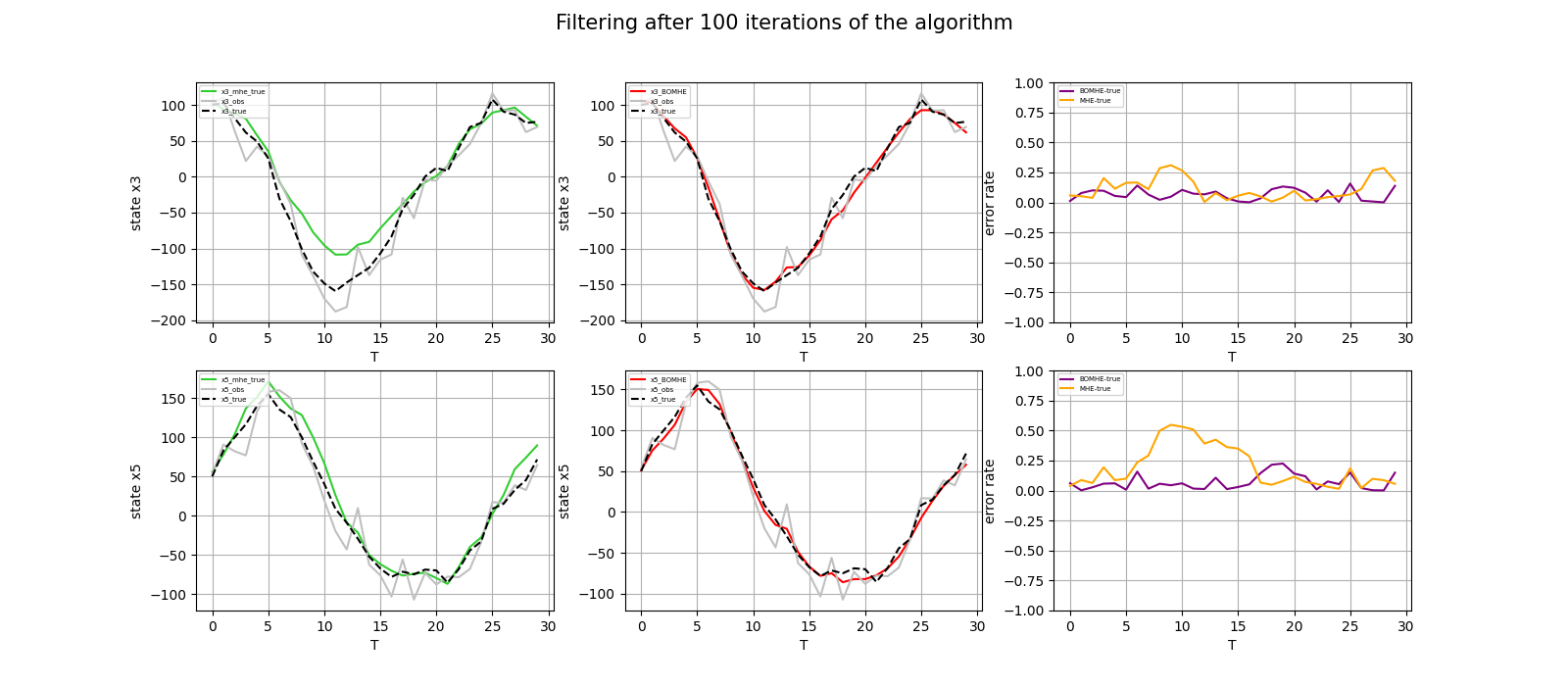}
\caption{It can be seen from the figure that the filter curve obtained by the BOMHE algorithm can track the actual system well. The exact linearized approximation is given here for the actual system When the system changes are more complex,  and the uncertainty of the system increases,  the tracking effect of the linearized system performance will be more affected.}
\label{fig:4}
\end{figure*}

\begin{figure}[htbp]
\centering
\includegraphics[width=\columnwidth]{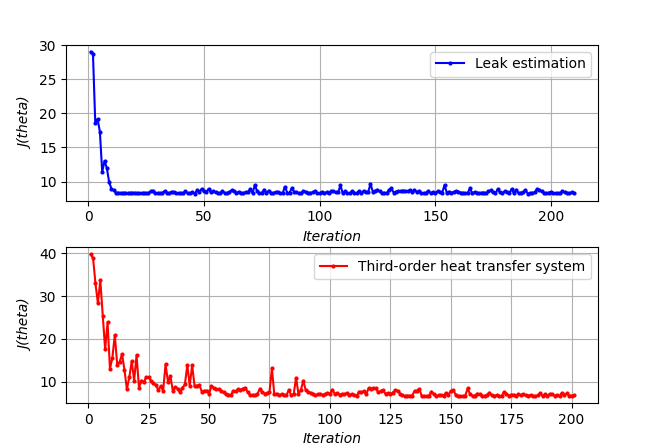}
\caption{The Iterative Convergence of Two Systems\centering}
\label{fig:3}
\end{figure}

These equations express the dynamics of the heat conduction system in state-space form, considering temperature, time, and thermal energy as state variables, and taking into account the influences of external input $u$ and ambient temperature $T_{env}$.
In the given system, \(x_1\) represents the system's temperature, \(x_2\) denotes time, and \(x_3\) signifies thermal energy. \(T\) is considered a constant ambient temperature, while \(u\) stands for external input. The parameters \(k_1\), \(k_2\), \(k_3\), and \(k_u\) define characteristics of the system. In this context, a linear observer is employed.$x:=(x_1,x_2,x_3)$ is the state vector.

In this context, we primarily focus on \(x_1\) and \(x_3\). The system observation vector is denoted as \(y\), and the system process noise is modeled as a normally distributed random variable with a covariance matrix \(Q\). The measurement noise is also assumed to be a normally distributed random variable with a covariance matrix \(R\). It is commonly assumed that the cost function's penalty coefficient matrix. Assume that the state vector \(x\) has an initial value of \([50, 0, 100]\). All observations are taken within the observation period \(T = 100\).

For this known nonlinear system,  the general practice in MHE is to do a first-order Taylor expansion at each state estimation point to get the Jacobi matrix of the system at that point as the linearization parameter of the system,  which has the drawback that the linearized system may lose the hidden information in the original higher-order terms and increases the computational burden. We choose to use BOMHE to select the parameter $\theta$ to linearize it,  which can omit the burden of calculating the Jacobi matrix for each estimation. Using a single linearized approximation to replace the original nonlinear system will have a significant error,  and the filtering effect is unsatisfactory. However,  as in the first example,  we aim to find the linearized system in the observation period to find the linearized approximation that is closest to the performance exhibited by the actual system.

\begin{table}[!t]
\setlength{\abovecaptionskip}{0cm}
\setlength{\belowcaptionskip}{0.3cm} %段后
\caption{Third-order heat transfer system\label{tab:table2}}
\centering
\begin{tabular}{|c|c|c|}
\hline
ALGORITHM& MHE-TRUE &  BOMHE\\
\hline
MAE & 26.189 & 20.856\\
\hline
\end{tabular}
\end{table}

Again,  this nonlinear system has two state values because considering the absence of a priori knowledge of the system dynamics model,  the parameter $\theta$ is still chosen randomly when initializing the GP,  where all values of the parameter $\theta$ are bounded by [-2, 2]. In evaluating this nonlinear system,  as in the first example,  the extra iteration steps are unnecessary. We use the same approach to give the values of MHE-TRUE and BOMHE,  where we can see that the filtering obtained by BOMHE performs better than the filtering obtained by linearizing at each sampling point when the nonlinearity of the system is high. Figure 3 performs the system simulated by the algorithm on the MHE at the 190rd iteration of the algorithm.
\subsection{Convergence and robustness}\label{sec:Convergence and robustness}
The convergence behavior of the MHE loss function over 200 iterations is presented in Figure 4. It can be observed that for leak detection, the loss function exhibits a rapid initial descent, reaching a relatively stable level after approximately 10 iterations. However, in the case of nonlinear models, it takes around 50 iterations to reach a relatively stable level. Moreover, to prevent convergence to local optima, Bayesian optimization actively explores unsampled points, resulting in slight fluctuations in the curve. Nevertheless, after 100 iterations, the loss function demonstrates convergence characteristics.
The robustness of the system is inherently satisfied as stated in the problem description, where the statistical properties of the disturbance noise for the system to be estimated are also unknown. Our objective is to maximize the filtering performance, rather than focusing on the precise characterization of the system dynamics. Therefore, there is no need to emphasize the accuracy of the system dynamics description. When the system is subjected to new disturbances, the algorithm will obtain a new approximate linearized model to maximize the filtering performance.

\section{Conclusion}\label{sec:Conclusion}
In this paper,  we mainly propose the BOMHE algorithm,  whose primary role is to perform MHE for an unknown system and estimate the system's actual performance from the observations. The method is proven to perform well in both linear and nonlinear systems through numerical simulations. In this paper,  to better illustrate the performance of the BOMHE algorithm on actual systems,  the dynamics of actual systems are given in the simulations. However,  the scenarios to which the algorithm applies usually do not include the a priori information of the dynamical system,  which means that the algorithm mainly applies to actual systems that are a complete black box model. In this paper,  the Bayesian parameters of the BOMHE algorithm are not selected. The acquisition function affects the final optimization performance to a large extent in black-box optimization problems,  and the choice of such parameters is a direction that needs attention. Since the algorithm does not have to consider the prior distribution of the system deterministically,  it may be more flexible in dealing with noise,  which allows us to consider the filtering performance under the influence of different noises,  not only Gaussian white noise.

%Bibliography
\bibliographystyle{unsrt}  
\bibliography{references}

\begin{thebibliography}{10}

\bibitem{rawlings2006particle}
James~B Rawlings and Bhavik~R Bakshi.
\newblock Particle filtering and moving horizon estimation.
\newblock {\em Computers \& chemical engineering}, 30(10-12):1529--1541, 2006.

\bibitem{alessandri2010advances}
Angelo Alessandri, Marco Baglietto, Giorgio Battistelli, and Victor Zavala.
\newblock Advances in moving horizon estimation for nonlinear systems.
\newblock In {\em 49th IEEE Conference on Decision and Control (CDC)}, pages 5681--5688. IEEE, 2010.

\bibitem{battistelli2018distributed}
Giorgio Battistelli.
\newblock Distributed moving-horizon estimation with arrival-cost consensus.
\newblock {\em IEEE Transactions on Automatic Control}, 64(8):3316--3323, 2018.

\bibitem{allan2019moving}
Douglas~A Allan and James~B Rawlings.
\newblock Moving horizon estimation.
\newblock {\em Handbook of model predictive control}, pages 99--124, 2019.

\bibitem{schiller2023lyapunov}
Julian~D Schiller, Simon Muntwiler, Johannes K{\"o}hler, Melanie~N Zeilinger, and Matthias~A M{\"u}ller.
\newblock A lyapunov function for robust stability of moving horizon estimation.
\newblock {\em IEEE Transactions on Automatic Control}, 2023.

\bibitem{simon2006optimal}
Dan Simon.
\newblock {\em Optimal state estimation: Kalman, H infinity, and nonlinear approaches}.
\newblock John Wiley \& Sons, 2006.

\bibitem{khuri2010response}
Andr{\'e}~I Khuri and Siuli Mukhopadhyay.
\newblock Response surface methodology.
\newblock {\em Wiley Interdisciplinary Reviews: Computational Statistics}, 2(2):128--149, 2010.

\bibitem{calandra2016bayesian}
Roberto Calandra, Andr{\'e} Seyfarth, Jan Peters, and Marc~Peter Deisenroth.
\newblock Bayesian optimization for learning gaits under uncertainty.
\newblock {\em Annals of Mathematics and Artificial Intelligence}, 76(1):5--23, 2016.

\bibitem{rokonuzzaman2021customisable}
Mohammad Rokonuzzaman, Navid Mohajer, Shady Mohamed, and Saeid Nahavandi.
\newblock A customisable longitudinal controller of autonomous vehicle using data-driven mpc.
\newblock In {\em 2021 IEEE International Conference on Systems, Man, and Cybernetics (SMC)}, pages 1367--1373. IEEE, 2021.

\bibitem{piga2019performance}
Dario Piga, Marco Forgione, Simone Formentin, and Alberto Bemporad.
\newblock Performance-oriented model learning for data-driven mpc design.
\newblock {\em IEEE control systems letters}, 3(3):577--582, 2019.

\bibitem{marco2016automatic}
Alonso Marco, Philipp Hennig, Jeannette Bohg, Stefan Schaal, and Sebastian Trimpe.
\newblock Automatic lqr tuning based on gaussian process global optimization.
\newblock In {\em 2016 IEEE international conference on robotics and automation (ICRA)}, pages 270--277. IEEE, 2016.

\bibitem{trimpe2014self}
Sebastian Trimpe, Alexander Millane, Simon Doessegger, and Raffaello D'Andrea.
\newblock A self-tuning lqr approach demonstrated on an inverted pendulum.
\newblock {\em IFAC Proceedings Volumes}, 47(3):11281--11287, 2014.

\bibitem{sorourifar2021data}
Farshud Sorourifar, Georgios Makrygirgos, Ali Mesbah, and Joel~A Paulson.
\newblock A data-driven automatic tuning method for mpc under uncertainty using constrained bayesian optimization.
\newblock {\em IFAC-PapersOnLine}, 54(3):243--250, 2021.

\bibitem{bansal2017goal}
Somil Bansal, Roberto Calandra, Ted Xiao, Sergey Levine, and Claire~J Tomlin.
\newblock Goal-driven dynamics learning via bayesian optimization.
\newblock In {\em 2017 IEEE 56th Annual Conference on Decision and Control (CDC)}, pages 5168--5173. IEEE, 2017.

\bibitem{muntwiler2022learning}
Simon Muntwiler, Kim~P Wabersich, and Melanie~N Zeilinger.
\newblock Learning-based moving horizon estimation through differentiable convex optimization layers.
\newblock In {\em Learning for Dynamics and Control Conference}, pages 153--165. PMLR, 2022.

\bibitem{jin2021new}
Xue-Bo Jin, Ruben~Jonhson Robert~Jeremiah, Ting-Li Su, Yu-Ting Bai, and Jian-Lei Kong.
\newblock The new trend of state estimation: from model-driven to hybrid-driven methods.
\newblock {\em Sensors}, 21(6):2085, 2021.

\bibitem{alessandri2008moving}
Angelo Alessandri, Marco Baglietto, and Giorgio Battistelli.
\newblock Moving-horizon state estimation for nonlinear discrete-time systems: New stability results and approximation schemes.
\newblock {\em Automatica}, 44(7):1753--1765, 2008.

\bibitem{beckers2019closed}
Thomas Beckers, Somil Bansal, Claire~J Tomlin, and Sandra Hirche.
\newblock Closed-loop model selection for kernel-based models using bayesian optimization.
\newblock In {\em 2019 IEEE 58th Conference on Decision and Control (CDC)}, pages 828--834. IEEE, 2019.

\bibitem{chakrabarty2021safe}
Ankush Chakrabarty and Mouhacine Benosman.
\newblock Safe learning-based observers for unknown nonlinear systems using bayesian optimization.
\newblock {\em Automatica}, 133:109860, 2021.

\bibitem{lederer2019uniform}
Armin Lederer, Jonas Umlauft, and Sandra Hirche.
\newblock Uniform error bounds for gaussian process regression with application to safe control.
\newblock {\em Advances in Neural Information Processing Systems}, 32, 2019.

\bibitem{seeger2004gaussian}
Matthias Seeger.
\newblock Gaussian processes for machine learning.
\newblock {\em International journal of neural systems}, 14(02):69--106, 2004.

\bibitem{klenske2015gaussian}
Edgar~D Klenske, Melanie~N Zeilinger, Bernhard Sch{\"o}lkopf, and Philipp Hennig.
\newblock Gaussian process-based predictive control for periodic error correction.
\newblock {\em IEEE Transactions on Control Systems Technology}, 24(1):110--121, 2015.

\bibitem{osborne2009gaussian}
Michael~A Osborne, Roman Garnett, and Stephen~J Roberts.
\newblock Gaussian processes for global optimization.
\newblock In {\em 3rd international conference on learning and intelligent optimization (LION3)}, pages 1--15. Citeseer, 2009.

\bibitem{shahriari2015taking}
Bobak Shahriari, Kevin Swersky, Ziyu Wang, Ryan~P Adams, and Nando De~Freitas.
\newblock Taking the human out of the loop: A review of bayesian optimization.
\newblock {\em Proceedings of the IEEE}, 104(1):148--175, 2015.

\bibitem{rao2002constrained}
Christopher~V Rao and James~B Rawlings.
\newblock Constrained process monitoring: Moving-horizon approach.
\newblock {\em AIChE journal}, 48(1):97--109, 2002.

\end{thebibliography}

\end{document}